\documentclass[prd,12pt,aps,superscriptaddress,preprintnumbers,showpacs,tightenlines,floatfix,nofootinbib,amssymb]{revtex4}
\usepackage{epsfig}

\newcommand{\beqn}{\begin{eqnarray}}
\newcommand{\eeqn}{\end{eqnarray}}

\newcommand{\eq}[1]{(\ref{#1})}
\newcommand{\cC}{{\cal C}}

\newcommand{\vn}{{\mathbf n}}
\newcommand{\ve}{{\mathbf e}}
\newcommand{\vx}{{\mathbf x}}
\newcommand{\dd}{{\mathrm d}}
\newcommand{\Z}{{Z \!\!\! Z}}

\newcommand{\ITEP}{\affiliation{Institute of Theoretical and
Experimental Physics, B.Cheremushkinskaya 25, Moscow, 117259, Russia}}

\begin{document}

\title{Liquid crystal defects and confinement in Yang-Mills theory}

\author{M.N.~Chernodub}\ITEP

\preprint{ITEP-LAT/2005-13}

\begin{abstract}
We show that in the Landau gauge of the SU(2) Yang-Mills theory the residual global symmetry supports
existence of the topological vortices which resemble disclination defects in the nematic liquid crystals and
the Alice (half-quantum) vortices in the superfluid ${}^3$He in the A-phase. The theory also possesses
half-integer and integer-charged monopoles which are analogous to the point-like defects in the nematic crystal and
in the liquid helium. We argue that the deconfinement phase transition in the Yang-Mills theory in the Landau gauge
is associated with the proliferation of these vortices and/or monopoles. The
disorder caused by these defects is suggested to be responsible for the confinement of quarks in the low-temperature phase.
\end{abstract}


\date{July 21, 2005}

\maketitle

The confinement of color in Quantum Chromodynamics is one of the unsolved problems in the quantum field theory. Nowadays
the commonly accepted point of view is that confinement is encoded in the dynamics of gluon fields which are described by
the Yang-Mills (YM) theory. Popular approaches to the color confinement are based on the dual superconductor
mechanism~\cite{ref:dual:superconductor} and on the center vortex picture~\cite{ref:center:vortex}. These approaches
suggest that the confinement is caused by a particular dynamics of special gluon configurations called the Abelian monopoles and the
central vortices, respectively. Besides the different dimensionality of these configurations, the formulation of these confinement
mechanisms requires a gauge fixing of the non-Abelian symmetry which is also different for both approaches.

Recently, an interest to describe the confinement mechanism in the
Landau gauge in terms of the particular configurations of the
gluon fields has emerged~\cite{ref:suzuki,Chernodub:2005jh}. The
Landau gauge is very attractive because it is very well studied
both in perturbative and non-perturbative approaches. Moreover,
this gauge is color-symmetric and it can be well formulated (up to
the Gribov-copy problem with most of other gauges also have) both
in the perturbative regime and in the continuum space-time.

In Ref.~\cite{ref:suzuki} the dual Meissner effect caused by
magnetic displacement currents was observed numerically in the
Landau gauge. The dual Meissner effect leads to the confinement of quarks via
formation of the chromoelectric string. This string is similar the Abrikosov
vortex~\cite{ref:Abrikosov} in the ordinary superconductors.

Another approach~\cite{Chernodub:2005jh} utilizes
the spin--charge separation idea \cite{ref:slave-boson,ref:splitting:antti} which is
widely used in condensed matter physics to describe the behavior
of strongly correlated electrons in high-$T_c$ cuprate
superconductors~\cite{ref:highTc}.
In Ref.~\cite{Chernodub:2005jh} the YM theory in the Landau gauge
was re-formulated as a nematic liquid crystal in an internal
space-time. In this paper we observe a relation between the YM theory and a
nematic crystal without the use of spin-charge separation of the gluon field.

We show that the existence of a residual global
symmetry -- which remains intact after the conventional Landau gauge fixing -- leads to existence
of specific topological defects which are analogous to
the defects in the nematic liquid crystals and in the superfluid
${}^3$He. There are many examples of physical systems in which the
global symmetries play an essential role: ferromagnetics, liquid
He${}_3$ and He${}_4$, various liquid crystals {\it etc}.
We point out that the (residual) global gauge symmetry in the Landau gauge is
a symmetry which can not be simply ignored.

We are working with the pure SU(2) Yang-Mills theory in the Euclidean space-time.
The Landau gauge~\eq{eq:F} is defined as a minimization of the gauge fixing functional,
\beqn
\min_\Omega F[A^{\Omega}]\,,\qquad F[A] = \int \dd^4 x \, {\left[A^a_\mu(x)\right]}^2\,,
\label{eq:F}
\eeqn
over the gauge transformations $\Omega \in SO(3)_{\mathrm{gauge}}$. The corresponding local differential gauge condition
is $\partial_\mu A^a_\mu =0$. The condition~\eq{eq:F} fixes the
$SO(3)_{\mathrm{gauge}}$ {\it gauge} color freedom up to the $SO(3)_{\mathrm{global}}$ {\it global} color group,
\beqn
SO(3)_{\mathrm{gauge}} \to SO(3)_{\mathrm{global}}\,,
\eeqn
because the gauge-fixing functional~\eq{eq:F} is invariant under the global (co\-or\-di\-na\-te\--in\-de\-pen\-dent)
transformation,
\beqn
A^a_\mu(x) \to \Omega^{ab}_{\mathrm{gl}} \, A^b_\mu(x)\,,\qquad \Omega_{\mathrm{gl}} \in SO(3)_{\mathrm{global}}\,.
\label{eq:global:transform}
\eeqn
Since the gluon field does not transform under the action of the center $\Z_2$ of the gauge group $SU(2)$,
the gauge symmetry of the {\it pure} SU(2)  Yang-Mills theory is in fact $SO(3) \sim SU(2)/\Z_2$.

The vector gauge field $A_\mu$ transforms in the fundamental representation of the residual $SO(3)_{\mathrm{global}}$ group,
Eq.~\eq{eq:global:transform}. One can construct the composite "color-spin" field
\beqn
C^{ab}(x) = A^c_\mu (x) A^c_\mu(x)\cdot\delta^{ab} - A^a_\mu (x) A^b_\mu(x)\,,
\label{eq:def:C}
\eeqn
which is a scalar with respect to space-time rotations and a rank-2 tensor with respect to the global
color transformations\footnote{Note that a first intension
to construct the pure octet representations in the form $A^a_\mu A^b_\mu - \delta^{ab}\, A^c_\mu A^c_\mu/3$
instead of Eq.~\eq{eq:def:C} would make a mechanical analogy discussed below a bit obscure.}.
The field $C$ transforms in the adjoint representation of the residual $SO(3)_{\mathrm{global}}$ group,
\beqn
C(x) \to \Omega \, C(x) \, \Omega^T\,,
\label{eq:C:transformation}
\eeqn
where the superscript $T$ stands for the transposition in the color indices. By construction,
the field $C$ is symmetric, $C = C^T$.

At this point it is convenient to introduce a mechanical interpretation of the matrix $C^{ab}$.
The definition~\eq{eq:def:C} is similar to the moment of inertia tensor of a solid body composed of
mass-centers (labeled by the integer $n$) with the masses $m^n$ located at the positions ${\mathrm r}_n$:
\beqn
I^{ij} = \sum_n m_n (r^2_n \,\delta^{ij} - r_n^i r_n^j)\,.
\label{eq:I}
\eeqn
At each point $x$ of the space-time the matrix $C^{ab}$, Eq.~\eq{eq:def:C}, corresponds to the
inertia tensor of a "solid body" consisting of four ($\mu=1,\dots,4$ in the four-dimensional space-time)
mass-centers with equal "masses" $m_1=\dots m_4=1$. The gauge field $A^a_\mu$ plays a role of a coordinate
of the $\mu^{\mathrm{th}}$ mass-center. The color transformation~\eq{eq:C:transformation}
of the matrix $C$ corresponds to a transformation of the inertia tensor~\eq{eq:I} under a spatial rotation.

At each point $x$ of the space-time the matrix $C^{ab}(x)$ can be diagonalized with the
help of the $SO(3)$ transformation $\Theta(x)$,
\beqn
C^{ab}(x) = \Theta(x) \, {\mathrm{diag}}(c_1,c_2,c_3) \, \Theta^T(x)\,.
\eeqn
The eigenvalues $c_k$, $k=1,2,3$ can be interpreted as "moments of inertia" defined with respect
to the orthonormal "principal axes of inertia" ${\ve}_k$, $k=1,2,3$, respectively.
The axes $\ve_k$ are normalized eigenvectors of the "moment of inertia tensor" $C$.
The transformation $\Theta(x)$ relates the default basis in the color space with the basis of the principal axes
of inertia at the point $x$. The eigenvalues of the matrix $C$ are ordered, $c_1 \geqslant c_2 \geqslant c_3$, and
below we assume that these eigenvalues are not degenerate unless it is stated otherwise.

To pursue this mechanical analogy further, the tensor $C$ can be associated an "ellipsoid of inertia" in the color space.
The directions of the axes of this ellipsoid coincide with principal axes of inertia
$\ve_k$, $k=1,2,3$, while the axes lengths
are proportional to the inverse square roots of the corresponding moments of inertia.
The direction of the longest axis of the ellipsoid of inertia
-- corresponding to the lowest principal moment of inertia -- defines a unit
vector ${\mathrm n}(x) \equiv \ve_3(x)$ in the color space.

The vector ${\mathrm n}(x)$ is direction-less since the vectors
$\pm\vn$ correspond to the same ellipsoid of inertia. Indeed, the
ellipsoid is invariant under the transformations of the $D_2$
group which is a finite subgroup of the $SO(3)$ group of
rotations. The elements of $D_2$ are $\pi$-rotations about any two
principal axis of the ellipsoid. The product of these elements
generate the $\pi$-rotation about the third axis. The
$\pi$-rotation of the ellipsoid about either $\ve_1$ or $\ve_2$
axis leads to the flip $\vn \to -\vn$. Thus, the eigenvector
corresponding to the lowest eigenvalue of the composite symmetric
field~\eq{eq:def:C} defines an arrowless vector in the color space
$\vn$ - or, in other words, a projective plane RP(2) element -- for a given
configuration of the gauge fields $A^a_\mu$.

The arrowless feature of the vector ${\mathrm n}(x)$ allows to
formulate a liquid crystal-like structure in the YM fields. The
spin field $\vn$ may also be associated with the largest principal
axes of the axially symmetric molecules ({\it i.e.}, which is also
called the "director field") in nematic liquid
crystals~\footnote{Note that in the liquid crystals the global rotations of the
director are related to the rotations of the coordinate space,
while in the YM model they are independent. However, this
difference is inessential and it does not influence the topological and dynamical
statements made below.}. The ordinary nematic
crystals~\cite{ref:nematics:review,ref:deGennes} are liquids composed of rod-like
molecules which are randomly oriented in the isotropic
(high-temperature) phase. At low temperatures the systems goes
into the liquid crystal phase in which the "rods" tend to align
parallel to the direction $\vn_0$ in macroscopic volumes. The
molecules in liquid crystals do not have a positional order while
being orientationally ordered.

The rod-like molecule in the nematic is invariant under (i) the $\Z_2$ group consisting of the $\pi$-rotations
about any axis perpendicular to the longest axis of the molecule and (ii) the $SO(2)$ group of rotations
about the longest axis. Thus, the physical space of the axial molecule -- corresponding to the ellipsoid of inertia
defined by the color tensor~\eq{eq:def:C} -- is the coset
\beqn
G/H = SO(3)/(\Z_2 \times SO(2))\,.
\label{eq:coset}
\eeqn
Let us imagine for a moment that we have integrated out in the YM partition function (fixed to the Landau gauge) all the degrees
of freedom but the spins $\vn$. Then at each point of the space-time the physical group of the $\vn$-spins would become
exactly the coset group~\eq{eq:coset}. Thus, after all degrees of freedom but $\vn$ are integrated out,
the $D_2$ invariance group of the "ellipsoid of inertia" must be replaced by the $\Z_2 \times SO(2)$ group which leaves the
arrowless vector $\vn$ intact.

One should note that the coset group~\eq{eq:coset} does {\it not} correspond to the symmetry breaking $G\to H$ in the YM theory.
In the pure YM theory the direction of the vector field $\vn$ changes from point to point leading to the vanishing
vacuum expectation value, $\langle \vn \rangle = 0$ since the color symmetry is obviously not broken in this case.
Equation~\eq{eq:coset} defines the group, the non-unit elements of which make the physically distinguishable changes to the
director field $\vn$.

Despite the absence of the symmetry breaking, still one can define defects associated with
topological invariants of the arrowless vector $\vn$. These definitions do not depends on the dynamics
of the model, while they do depend on the possibility to define a unit vector with particular symmetry
properties~\eq{eq:coset}. The coincidence of the physical group~\eq{eq:coset} with the one of the nematic
crystals~\cite{ref:nematics:review,Michel:1980pc} allows us to
discuss topological (with respect to the residual $SO(3)_{\mathrm{global}}$ group) defects in the
Landau gauge of the YM theory using the corresponding condensed matter analogues.

The physically interesting are the first four homotopy groups of the physical space~\eq{eq:coset}. The $\pi_0$ group is trivial,
$\pi_0(G/H)=1$, telling us that there are no topologically stable domain walls made of the director field $\vn$.

\begin{figure}[!thb]
\begin{center}
\begin{tabular}{ccc}
\includegraphics[scale=0.4,clip=true]{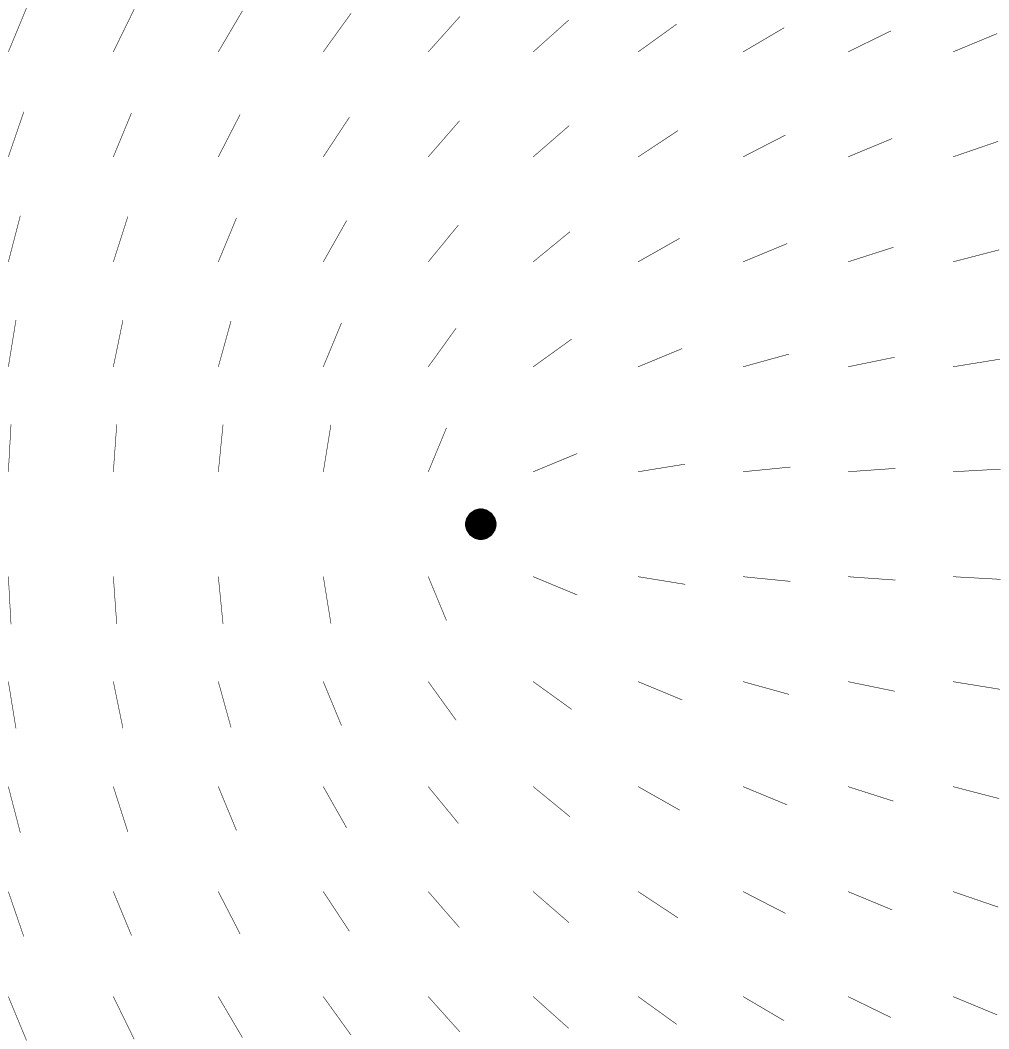} \hspace{5mm} &
\includegraphics[scale=0.4,clip=true]{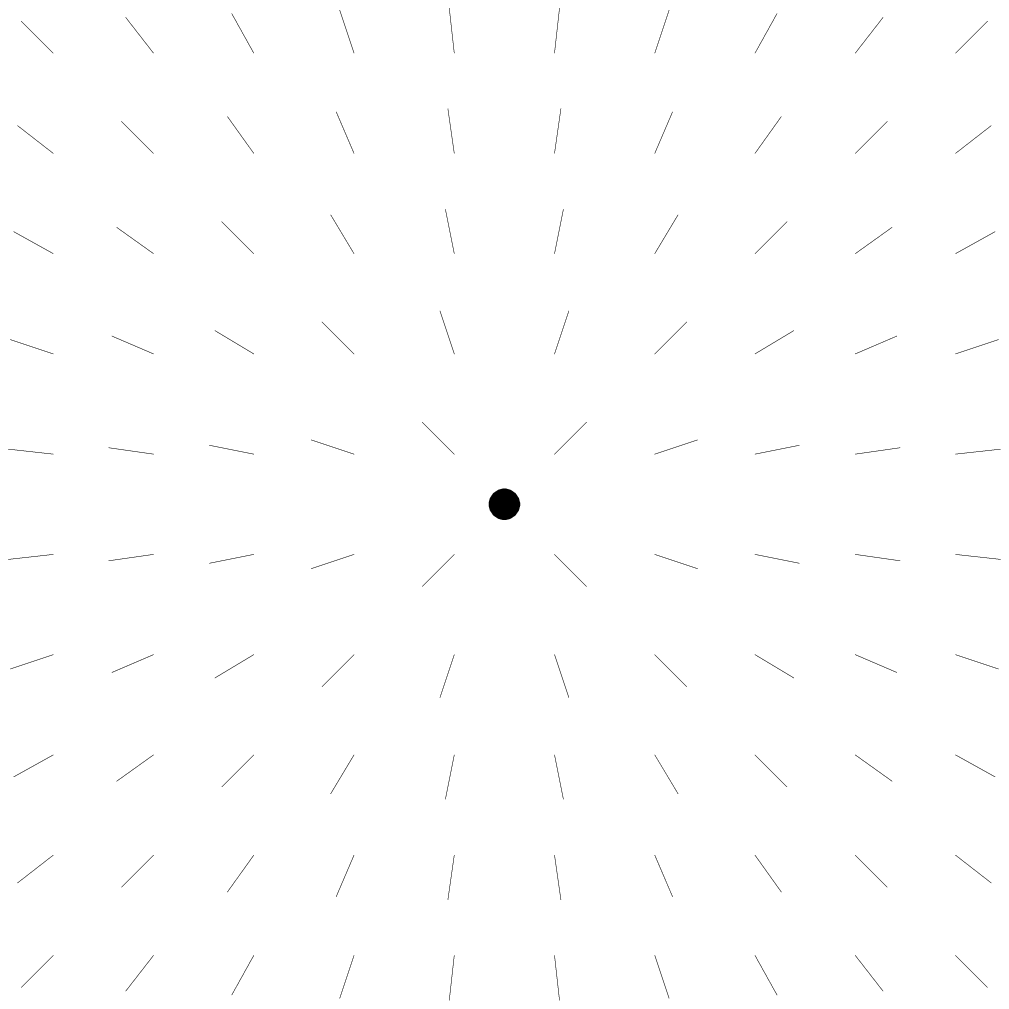} \hspace{5mm} &
\includegraphics[scale=0.4,clip=true]{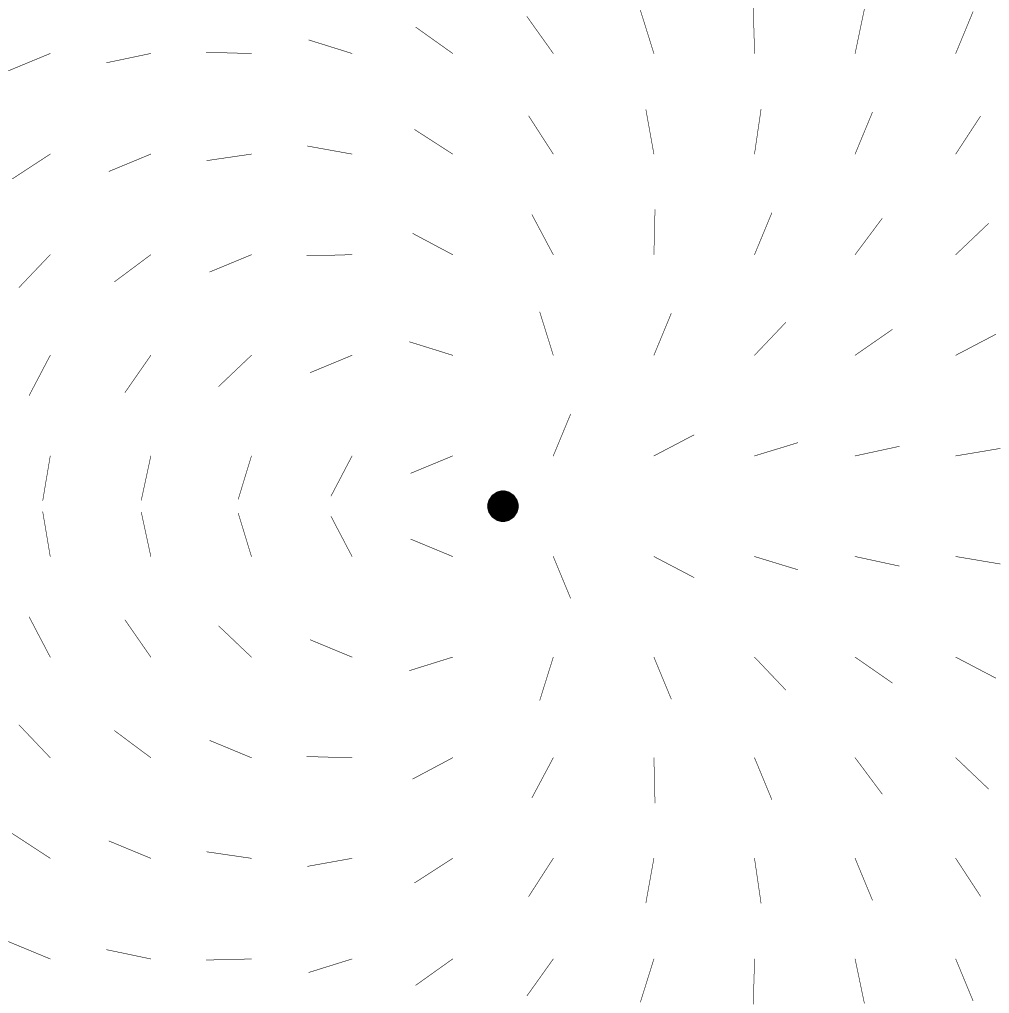} \\
(a) & (b) & (c)
\end{tabular}
\end{center}
\vspace{-4mm}
\caption{The schematic behavior of the vortex director field $\vn_{\mathrm{vort}}$, Eq.~\eq{eq:vortex},
in the color space. Vortices with (a) $n=1/2$ (topologically stable), (b) $n=1$ (unstable) and (c) $n=3/2$
(unstable) winding numbers are shown.}
\label{fig:vortex}
\end{figure}

The first homotopy group is nontrivial, $\pi_1(G/H)=\Z_2$, indicating that there are topologically stable vortices characterized
be the only nontrivial element of the $\Z_2$ group. The $\Z_2$-vortices are very well known as "disclination defects" in the
physics of liquid crystals~\cite{ref:nematics:review}. The typical example of the straight static vortex going perpendicularly to the
$x-y$ plane is described by the director field of the form:
\beqn
\vn_{\mathrm{vort}} = (\cos n \varphi,\sin n \varphi,0)\,
\label{eq:vortex}
\eeqn
where $\varphi$ is the azimuthal angle in the $x-y$ plane and
$n$ is the topological number which may take integer and half-integer values. The integer-values vortices are unstable, and
the topologically stable elementary vortex has the fractional circulation number $n=1/2$. The vortices with $n=1/2$, $n=1$
and $n=3/2$ are visualized in Figures~\ref{fig:vortex}(a),(b) and (c), respectively.
In the center of the vortex the color tensor~\eq{eq:def:C} must be at last double degenerate to support the
two-dimensional hedgehog-like singularity~\eq{eq:vortex}.

The presence of the half-winding vortices is possible because the directions $\vn$ and $-\vn$ are physically indistinguishable.
As it can be seen from Figure~\ref{fig:vortex}(a) the director field $\vn$ rotates in $180^\circ$ degrees in the color space,
({\it i.e.}, $\vn \to -\vn$), as one makes the full $360^\circ$-turn around the vortex position. These vortices share
similarities with the central vortices observed numerically in the Maximal Center gauge~\cite{ref:center:vortex} of the YM theory,
as well as with the so-called Alice strings discussed both in the particle physics~\cite{Schwarz:1982ec} and
in the condensed matter physics ({\it i.e.}, as half-quantum vortices which exist both in nematic liquid
crystals~\cite{ref:nematics:review,ref:deGennes} and in A-phase of the superfluid ${}^3$He~\cite{ref:Volovik:Mineev,ref:Volovik:book}).

The non-triviality of the second homotopy group, $\pi_2(G/H)=\Z_+$ with $\Z_+ \equiv \Z/\Z_2 = 0,1,2,\dots$ guarantees
the presence of the topologically stable monopoles. The monopoles are particle-like objects characterized by positive
half-integer winding numbers $m=M/2$, $M \in \Z_+$, which count the number of sweeping of the spatial sphere into
the sphere of the color group. Due to the $\Z_2$--identification $\vn\to-\vn$, the number $m$ can be half-integer.

The monopole current can be formulated in terms of the winding number,
\beqn
k_\mu = \frac{1}{8\pi} \epsilon_{abc}\epsilon_{\mu\nu\alpha\beta}
\frac{\partial n^a}{\partial x_\nu} \frac{\partial n^b}{\partial x_\alpha} \frac{\partial n^c}{\partial x_\beta}\,.
\label{eq:k}
\eeqn
The current is represented as $\delta$-singularities on a set of closed loops $\cC$  parameterized by the four-vector $X(\tau)$,
\beqn
k_\mu(x) = \sum_{\cC} m_\cC\, \int_\cC \!\dd \tau\,
\frac{\partial X^\cC_\mu(\tau)}{\partial \tau} \, \delta^{(4)}(x - X^\cC(\tau))\,,
\label{eq:k:delta}
\eeqn
where $m_\cC$ is the charge ascribed to the loop $\cC$.
By construction~\eq{eq:k}, the monopole current is closed, $\partial_\mu k_\mu = 0$. A three-dimensional analogue of Eq.~\eq{eq:k}
was used in Ref.~\cite{Blaha:1975jk} to formulate $\pi_2$-singularities in the superfluid Helium and in the liquid crystals.
Note that at the center of the monopole the color tensor~\eq{eq:def:C} must be triple-degenerate to support the
three-dimensional hedgehog-like singularity~\eq{eq:monopole}.

Due to the identification  $\vn \leftrightarrow - \vn$ the monopole winding numbers $\pm m_\cC$ correspond to the same monopole
according to Eq.~\eq{eq:k}. This feature is perfectly consistent with the presence of the Alice $\Z_2$-vortices discussed
above, because the monopole with the charge $m$ transforms into the (physically the same!) monopole with the charge $-m$
after circling around the Alice string~\cite{ref:Volovik:book}.

A class of the static unit director fields corresponding to the $m$-charged monopoles
is given in Refs.~\cite{ref:saupe,Blaha:1975jk}:
\beqn
\vn_{\mathrm{mon}} = (\sin \vartheta(\theta,m) \cdot \cos m \varphi,
\sin \vartheta(\theta,m) \cdot \sin m \varphi, \cos m \varphi)\,,
\label{eq:monopole}
\eeqn
where $\theta$ and $\varphi$ are polar and azimuthal angles, and
$\vartheta(\theta,m) = 2\arctan \left\{\left[\tan(\theta/2)\right]^m\right\}$.
The positive topological number $m$ may take integer and half-integer values. The current~\eq{eq:k:delta}
corresponding to the configuration~\eq{eq:monopole} can be calculated with the help of Eq.~\eq{eq:k},
$k_\mu({\mathbf{x}},t) = m\,\delta_{\mu,4}\,\delta^{(3)}(\vx)$. Note that the director field for the $m=1$ monopole
is just the standard hedgehog, $\vn_{\mathrm{mon}} = \vx$.

\begin{figure}[!thb]
\begin{center}
\begin{tabular}{ccc}
\includegraphics[scale=0.4,clip=true]{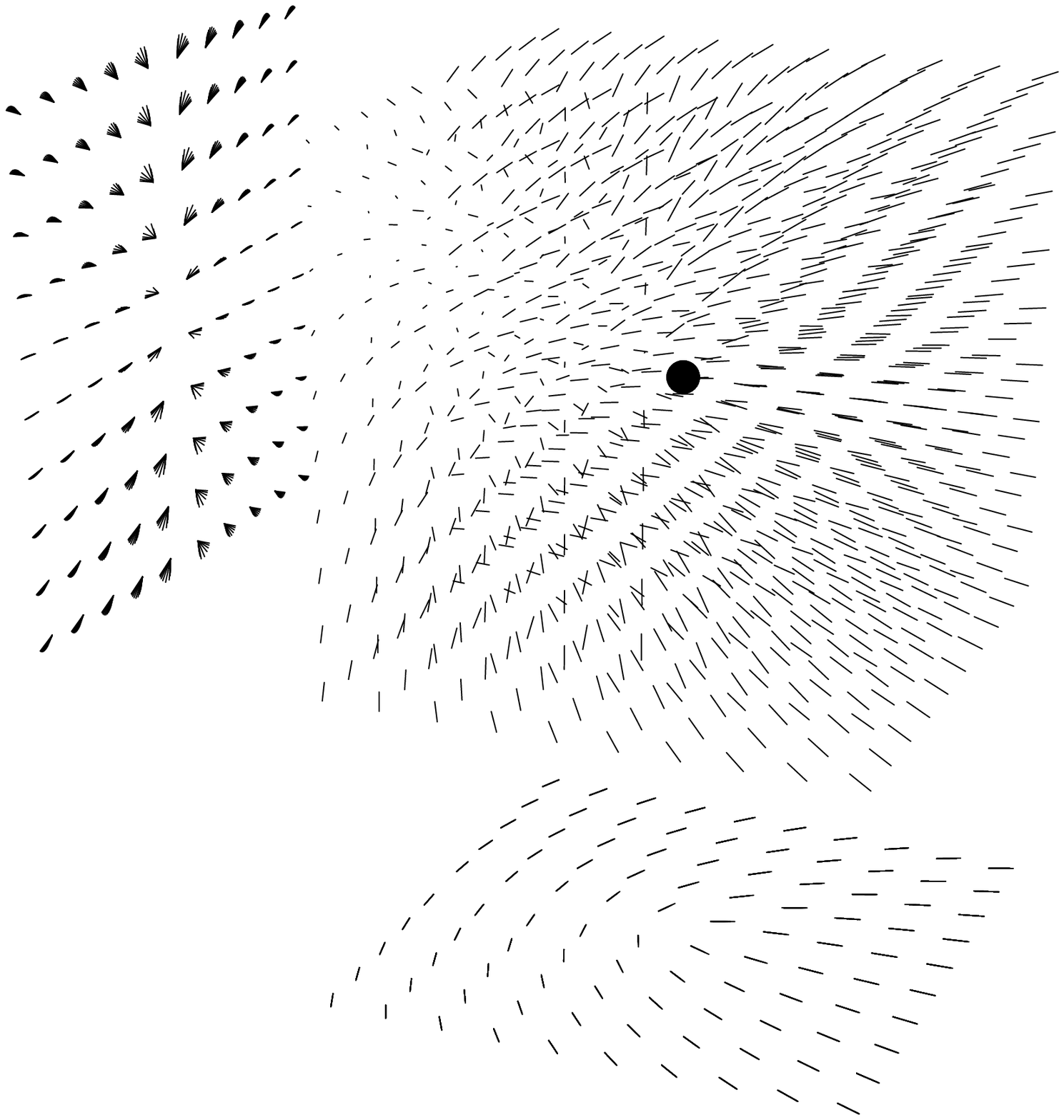} \hspace{5mm} &
\includegraphics[scale=0.4,clip=true]{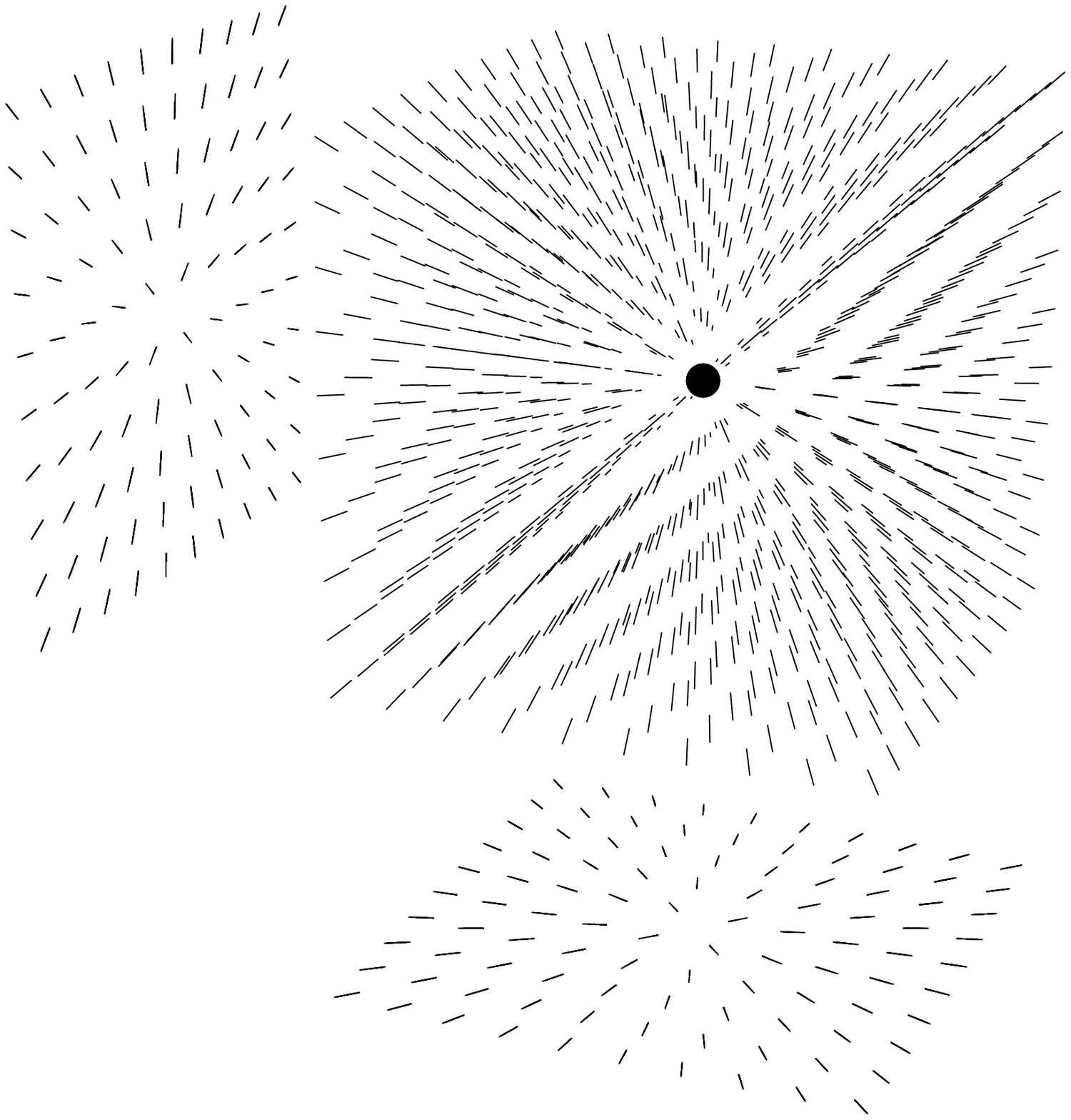} \\
(a) & (b)
\end{tabular}
\end{center}
\vspace{-4mm}
\caption{The same as in Fig.~\ref{fig:vortex} but for monopoles~\eq{eq:monopole} with the winding numbers
(a) $m=1/2$ and (b) $m=1$. The projections of the director field~$\vn$ into two planes are also shown.}
\label{fig:monopole}
\end{figure}

The examples of the static $m=1/2$ and $m=1$ monopoles~\eq{eq:monopole} are visualized in a spatial time-slice
in Figures~\ref{fig:monopole}(a), (b), respectively.
The bottom projection of the $m=1/2$ monopole, Figure~\ref{fig:monopole}(a), shows a shadow of the attached
$n=1/2$ vortex, Figure~\ref{fig:vortex}(a).

The third homotopy group is also non-trivial, $\pi_3(G/H)=\Z$, which leads to the appearance of the
instantons in the four dimensional sense.

As an example of a particular configuration of the gluon fields, consider the BPS monopole
solution~\cite{ref:BPS} to the SU(2) Yang--Mills equation of motion:
\beqn
A^a_i & = & \frac{1}{g}\, f(r) \varepsilon_{iab} {\hat x}^b\,,\qquad f(r) = \frac{1}{r} \Bigl(1 - \frac{r}{\sinh r}\Bigr)\,,
\label{eq:BPS:spatial}\\
A^a_4 & = & \frac{1}{g}\, h(r) \, {\hat x}^a\,,\qquad \hspace{5mm} h(r) = \frac{1}{r} \Bigl(r \coth r -1 \Bigr)\,,
\label{eq:BPS:temporal}
\eeqn
where ${\hat x}^a = x^a/|\vx|$, and $r \equiv |x|$ is assumed to be scaled by an arbitrary factor, $r \to r_0 \cdot r$,
to make a dimensionless quantity. This static configuration satisfies the local Landau gauge condition, $\partial_\mu A^a_\mu =0$.

The color tensor $C^{an}$, Eq.~\eq{eq:def:C}, evaluated at the monopole configuration~(\ref{eq:BPS:spatial},\ref{eq:BPS:temporal})
has three eigenvalues, $c_1=c_2 = (f^2(r)+h^2(r))/g^2$ and $c_3 = 2 f^2(r)/g^2$. Since the two out
of the three eigenvalues are degenerate, the "ellipsoid of inertia" is axially symmetric. Moreover, $f(r)<h(r)$ for $r>0$,
therefore $c_1(r)=c_2(r) > c_3(r)$, and the ellipsoid is always rod-like (contrasted to a disk-like form). The longest axis
of the ellipsoid forms a hedgehog-like structure around the center of the monopole, $\vn \equiv \ve_3 = \hat\vx$.
Thus, in the Landau gauge the BPS YM-monopole can be identified with the $m=1$ charge global monopole~\eq{eq:monopole}, which
can be called as "arrowless hedgehog" due to the $\Z_2$ symmetry $\vn \leftrightarrow -\vn$. Note that in the center of the monopole
the tensor~\eq{eq:def:C} is triple-degenerate, $c_1(0)=c_2(0)=c_3(0)=0$.

Before discussing the dynamics of the liquid crystal defects in the YM theory let us mention an attempt~\cite{ref:Fedor} to
construct an effective nematic theory from the YM theory in an explicitly gauge-invariant form.
The nematic variables were constructed from the field-strength tensor of the SU(2) gauge field.
The numerical results~\cite{ref:Fedor} have revealed a dominance of the ultraviolet fluctuations in the dynamics of the
SU(2) gauge-invariant nematic variables
both in confinement (low temperature) and in the deconfinement (high temperature) phases. We believe that the
gauge-fixing used in our approach smoothes the gauge fields making the nematic field less random.

Since the color symmetry is not broken in the YM theory the director field does not
have a constant direction in the color space, $\langle \vn \rangle = 0$.  Therefore both the confinement and the deconfinement
phases correspond -- in the language of the nematic liquid crystals -- to the (color) isotropic phases, where the density of the
topological defects is non-zero due to large-angle fluctuations of the director field.

In the confinement phase a network of the defects is
likely to be percolating so that any two infinitely-separated
points have a finite probability to be connected by a
world-trajectory of the defects. Generally, the percolation of a defect trajectory
means an existence of a non-vanishing low-momentum (infrared) component of a corresponding field,
which implies the condensation of the defect.
The percolation (condensation) feature is shared by many
realistic theories possessing the topological defects. The
percolation happens in the phases where the symmetry group
(responsible for the appearance these defects) is unbroken. The
relevant examples are the symmetric phases of the Abelian
Higgs (or, the Ginzburg-Landau) model~\cite{ref:Higgs:Abelian} and the
non-Abelian Higgs (or, the Standard Electroweak) model~\cite{ref:Higgs:EW} where
the monopole-like and vortex-like defects are known to be
proliferating at infinitely large distances. Note that in nematic crystals
the monopole density (contrary to the vortex density) is expected to be low due to
a topological argument of Ref.~\cite{ref:Hindmarsh}. Nevertheless, a low-density monopole
network can also be percolating.

To discuss the behavior of the monopoles in the deconfinement
phase we note, that both integer and half-integer monopoles in the
Landau gauge are characterized by the specific
(hedgehog-like, as in the examples shown in Figure~\eq{fig:monopole})
behavior of the director field in a local
three-dimensional time-slice perpendicular to the direction of the
monopole current~\eq{eq:k:delta}. In order to support a non-zero
sphere-to-sphere winding number~\eq{eq:k} the director field must
evolve in all three dimensions in the local vicinity of the
monopole. The hedgehog structures involving the Euclidean time
direction are suppressed at high temperatures by high Matsubara
frequencies, leading to a suppression of the monopoles moving in
spatial directions.

Thus, at high temperatures the monopoles
must be static, and, as a consequence, non-percolating. The
high and the low temperature regimes must be separated by a
percolation transition, which is very likely to at the same
temperature as the deconfinement phase transition as it happens in
the the Abelian and non-Abelian Higgs theories~\cite{ref:Higgs:Abelian,ref:Higgs:EW}. A similar
property is also observed numerically for the Abelian monopoles in the Maximal
Abelian gauge of the Yang-Mills theory~\cite{ref:dual:superconductor:percolation}.

Similar arguments can be applied to the half-quantum vortices in the Landau gauge: they are expected to be proliferating in
the confinement phase and tend to be static in the deconfinement phase. Since the vortices are line-like defects, two different
possibilities can be in the
deconfinement phase: (i) the vortices may exist in the form of dominantly static and non-percolating short loops, (ii)
or the static vortices may exist in the form of long loops leading to a spatial percolation of the vortex trajectories.
The experience gained with the center vortices in the Maximal Center gauge~\cite{ref:Center:percolation} tells us that it is
the first option that is likely to be realized in the high-temperature phase.

The dynamics of the topological defects is important because it is related to the confinement of color.
In the confinement phase the percolating defects should cause a disorder which should lead to the area law of
sufficiently large Wilson loops. In the deconfinement phase the defects
are expected to be dominantly static and the disorder spreads only in spatial
dimensions leading to the deconfinement of the static quarks and to confinement
of the "spatial" quarks in a sense of the area-law for the spatial Wilson loops.

Let us discuss the (dis)order parameter phase transition within the
described (nematic) picture. In the language of the topological
defects, the disorder parameter for the deconfinement phase
transition is given by the percolation probability for the
monopoles (and, probably, vortices) to proliferate for infinitely
long distances. This probability should be zero in the
deconfinement phase and non-zero in the confinement phase.

In the language of fields a best candidate to discuss the transition might be the color
tensor $C^{ab}$, Eq.~\eq{eq:def:C}, because the field $C^{ab}$ (up to an inessential color-singlet part) looks
similar to the diamagnetic susceptibility in the nematic liquid crystals. The diamagnetic susceptibility is known
to be the order parameter of the nematic-isotropic phase transition in the liquid crystals~\cite{ref:nematics:review}.
In the nematic (low temperature) phase the director field is dominantly constant leading to
an anisotropy in the diamagnetic susceptibility, while in the high temperature phase the director field is random and
the diamagnetic susceptibility is isotropic.
Since the color symmetry is unbroken in the YM theory, one must
have $\langle C^{ab} \rangle \propto \delta^{ab}$ both in the confinement and in the deconfinement phases.
Consequently, the field $C^{ab}$ can not be used as an order parameter for the deconfinement phase transition.

Thus, the deconfinement phase transition in the YM must be
associated with a transition from one isotropic phase of the nematic crystal to another
isotropic phase. The two distinct isotropic phases
were indeed observed in the lattice numerical simulations of the three-dimensional nematic
crystals~\cite{ref:phase}. One of such isotropic phases is characterized the defect condensation.
In the language of the YM theory it corresponds to the confinement phase. Another isotropic phase
-- called the topologically ordered phase~\cite{ref:phase} -- is characterized by the absence of the condensate
(the analog of the deconfinement phase in the YM theory). Surprisingly, the isotropic-to-isotropic
transition in the nematic crystal is the second-order phase
transition lying in the 3D Ising universality class~\cite{ref:phase}.
The fact that the order and the universality class of the phase transition in the nematic crystal coincide with the
order and the class of the finite-temperature phase transition in the SU(2) YM theory~\cite{ref:universality}
provides an an additional support in favor of the liquid crystal interpretation of the YM theory in the Landau gauge.

\begin{acknowledgments}
The author is supported by grants RFBR 04-02-16079, RFBR 05-02-16306, DFG grant 436 RUS 113/739/0 and MK-4019.2004.2.
The author is grateful to F.V.~Gubarev for useful discussions, and to A.~Niemi for turning the attention of the author to
Ref.~\cite{ref:Volovik:book}. The author is also thankful for the kind hospitality and stimulating environment
to the members of the Theoretical Particle Physics group of Humboldt University (Berlin) where a part of this work was done.
\end{acknowledgments}

\end{document}